\documentclass{nature}

\usepackage{amsmath,amssymb,amsfonts}
\usepackage{cite}
\usepackage{CJK}
\usepackage{bm}
\usepackage{tikz}
\usepackage{pgfplots}
\usetikzlibrary{matrix,fit}
\usepackage{graphicx} 
\usepackage{braket}
\usepackage{subfig}
\usepackage{caption}
\usepackage{mathrsfs}
\usepackage{float}
\usepackage[utf8]{inputenc}
\usepackage{pgflibraryarrows}
\usepackage{lipsum} 
\usepackage{pgflibrarysnakes}
\usepackage{authblk}
\usepackage{lineno}

\usepackage[labelsep=none]{caption}
\usepackage[labelfont=bf]{caption}
\usepackage[absolute,overlay]{textpos}

\title{Higher-order Quantum Spin Hall Effect of Light}

\makeatletter
\renewcommand*{\@fnsymbol}[1]{\ensuremath{\ifcase#1\or \dagger\or *\or \ddagger\or
    \mathsection\or \mathparagraph\or \|\or **\or \dagger\dagger
    \or \ddagger\ddagger \else\@ctrerr\fi}}
\makeatother

\author{Biye Xie$^{1,3}$\thanks{These authors contributed equally to
    this work.} ,
  Guangxu Su$^{1,2 \dagger}$, Hong-Fei Wang$^{1,3\dagger}$, Feng Liu$^{4}$, Lumang Hu$^{1,2}$ Si-Yuan
  Yu$^{1,3}$, Peng Zhan$^{1,2}$\thanks{Corresponding authors: zhanpeng@nju.edu.cn, luminghui@nju.edu.cn., zlwang@nju.edu.cn and yfchen@nju.edu.cn},
  Ming-Hui Lu$^{1,3,5 *}$, Zhenlin Wang$^{1,2 *}$, \& Yan-Feng Chen$^{1,3 *}$}
\usepackage{graphicx}
\makeatletter
\let\saved@includegraphics\includegraphics
\AtBeginDocument{\let\includegraphics\saved@includegraphics}
\renewenvironment*{figure}{\@float{figure}}{\end@float}
\makeatother

\begin{document}
\maketitle

\begin{affiliations}
 \item National Laboratory of Solid State Microstructures, Collaborative Innovation Center of Advanced Microstructures, Nanjing University, Nanjing 210093, China
 \item School of Physics, Nanjing University, Nanjing 210093, China
 \item Department of Materials Science and Engineering, Nanjing University, Nanjing 210093, China
 \item Department of Nanotechnology for Sustainable Energy, School of
   Science and Technology, Kwansei Gakuin University, Sanda 6691337, Japan
\item Jiangsu Key Laboratory of Artificial Functional Materials, Nanjing 210093, China

 \thanks
\end{affiliations}
\newpage

\begin{abstract}
Band topology and related spin (or pseudo-spin) physics of photons provide us
with a new dimension for manipulating light, which is potentially useful for
information communication and data storage~\cite{spin1, spin2,spin3,SHE1,SHE2}.
Especially, the quantum spin Hall effect of light, where
electromagnetic waves propagate along surfaces of samples
with strong spin-momentum locking, paves
the way for achieving topologically protected photonic spin transport~\cite{QSHL}.
Recently, the conventional bulk-edge correspondence of the band topology has been extended
to higher-order cases that enables the explorations of topological states with
codimensions larger than 1 such as hinge and corner states~\cite{HOTI1,HOTI2,HOTI3,HOTI4,HOTI5,HOTI6,HOTI9,HOTI10,HOTI11,HOTI12,
HOTI13, BIYEPRL,DJWPRL}. Here, for the first time, we demonstrate a higher-order quantum spin Hall effect of light by utilizing an all-dielectric $C_{6v}$-symmetric
photonic crystal. We observe corner states with
opposite pseudospin polarizations at different corners owing to nontrivial
higher-order topology and finite spin-spin coupling~\cite{Liu}. By
applying the spin-polarized excitation sources, we can selectively
excite the corner states at different spatial positions through
spin-momentum-locked decaying edge states, resembling the quantum spin Hall effect in a higher-order manner.
Our work which breaks the barriers between the spin photonics and higher-order topology opens the frontiers for studying lower-dimensional spinful classical surface waves and supports explorations in robust communications.

\end{abstract}

Trapping and guiding the flow of light lie at the heart of modern
integrated photonic devices that are crucial for the realizations of
photonic quantum communication and computing~\cite{Integrated,Quantum1,Quantum2}.
However, owing to unavoidable disorders and imperfections during
the fabrication process, the localization and propagation of light in
traditional photonic devices suffer from fragility and
backscattering.
Fortunately, by invoking the spin (or pseudospin) degree of freedom of
photons combined with the nontrivial band topology, we can enable spinful light to
propagate unidirectionally
without backscattering, which is now known as the
quantum spin Hall effect (QSHE) of light~\cite{Hu1, Hu2, QSHL}.
This effect which describes the boundary states of a sample is characterized by a nontrivial topological invariant and supports boundary spin (or pseudospin) transport, yielding the realization of photonic topological insulators and striking advances in spinful photonic devices~\cite{PTI1, PTI2, PTI3, PTI4, PTI5, PTI6, PTI01, PTI7, PTI9}.
However, the conventional QSHE of light only focuses on the one-dimensional propagation of light, and to realize the higher-order QSHE in which the light is localized at a designated position with both topological protection and spin
polarization is a far more challenging task.

Recently, higher-order topological insulators (HOTIs) have been introduced as a new kind of topological phases of matter which go beyond the conventional bulk-boundary correspondence~\cite{HOTI1,HOTI2, HOTI3, HOTI4, HOTI5, HOTI6, HOTI9, HOTI10,HOTI11, HOTI12, HOTI13, BIYEPRL, DJWPRL} . If we define the codimension $l$ of an $n$-dimensional state in an $m$D system as $l=m-n$, then an $l$th-order topological insulator can be defined as a topological insulator with $l$-codimensional boundary states. These lower-dimensional boundary states such as hinge states and corner states provide new degree of freedom to manipulate waves and support integrated topological devices. Previous experimental realizations of HOTIs are all limited to explore the existence and robustness of these boundary states~\cite{HOTI9, HOTI10, HOTI11, HOTI12, HOTI13,BIYEPRL}. The internal degrees of freedom of waves, such as the spin and pseudospin, offer new degrees of freedom to investigate the wave physics and
sustain vast applications from astronomy to quantum information
processing which have so far not been intertwined with the
higher-order topology.

In this article, we propose a spinful photonic second-order topological insulators with corner states. Moreover, we demonstrate that corner states are induced and separated from bulk and edge states by
the filling anomaly~\cite{HOTIC} in the $C_{6v}$ symmetric topological
crystalline insulators as shown in Fig. 1a.
Intriguingly, we show that the corner states in our implementation have opposite pseudospins
(denoted as spinful corner states)
which are realized by the linear combination of the $d$ states in the folded band structure as shown in Fig. 1b. 
As a consequence, when the pseudospin-pseudospin coupling is considered in a finite size structure, the
pseudospins will have nonzero polarizations~\cite{Liu}.
To observe these nonzero pseudospin polarizations, we experimentally fabricate a sample consist of a second-order photonic topological insualtor and a trivial topological insualtor as shown in Fig. 1c. We then apply a
pseudospin-dependent excitation source at the middle of the interface between two topologically distinct area and observe the near-field electromagnetic wave. We find that the wave are guided to and localized at different corners with opposite pseudospin polarization, demonstrating a higher-order QSHE in HOTIs as shown in Fig. 1c.

\textbf{Lattice structure and nontrivial higher-order topology}
  
The photonic crystal (PC)~\cite{PC1,PC2} we considered here is a triangular lattice
of hexagonal clusters with six dielectric rods in each cluster as
shown in the insets of Fig. 2a.
For simplicity, we only consider the transverse magnetic (TM) mode.
The distance among the dielectric rods determines the coupling
strength and hence there are two competing parameters: the inter-cell
couplings $t_{inter}$ and the intra-cell couplings $t_{intra}$ that
determine the band structure of the PC. If $t_{inter}=t_{intra}$,
the PCs has a honeycomb lattice and the band structure of the
triangular lattice is the band-folding version of the one in honeycomb
lattice (The Dirac cones at the Brillioun zone corners $K$ and
$K^{\prime}$ of honeycomb lattice are folded to the Brilloun zone
center $\Gamma$ of the triangular lattice)~\cite{Hu1,Liu}.
By adjusting the inter-cell couplings and intra-cell couplings,
the Dirac cone at $\Gamma$ can be gapped out and a topological phase
transition occurs when the PCs evolves from $t_{inter}<t_{intra}$
(denoted as the expanded lattice which is topologically non-trivial) to $t_{inter}>t_{intra}$ (denoted as
the shrunken lattice which is topologically trivial)[See section I of the Supplementary Information
(SI)~\cite{SUP}.].
In our study, we set $a=18.0mm$ $r=1.8mm$ $h=25.0mm$ $\epsilon=8.5$ as
lattice constant, radius of rods,
height of the rods and relative dielectric constant of the rods
respectively for all lattices.
$b_1=5.0mm$ and $b_2=6.8mm$ are the distances between the nearest-neighbour rods in the unit cell which are proportional to the inter-cell coupling strengths for the ordinary insulator (OI) and topological insulator (TI) respectively. 
(The intra-cell coupling strengths are automatically determined once the inter-cell coupling strengths are determined since the lattice is the tessellation of the unit cell)

For the expanded lattice case, the PC has a dimensional hierarchy of
the higher-order topology in which both the first- and the
second-order photonic topological insulator emerge~\cite{BIYEPRL}.
The first-order and the second-order topological insulating phase are
characterized by the bulk polarization $\textbf{P}=(P_x, P_y)$
and the secondary topological index~\cite{HOTIC} $Q^c$ respectively.
Based on the eigenvalues of the $C_6$ operation at high symmetric
points in the first Brillouin zone, these topological invariants are obtained as follows\cite{HOTIC}, 
\begin{equation}
\begin{aligned}
P_x=P_y &=[K_1]+2[K_2]=0 \quad mod \quad 1,\\
Q^c &=\frac{1}{4}[M_1]+\frac{1}{6}[K_1] \quad mod \quad1
\end{aligned}
\end{equation}
where $[\Pi_p]=\#\Pi_p-\#\Gamma_p$ and $\#\Pi_p$ is defined as the
number of bands below the band gap with rotation eigenvalues
$\Pi_p=e^{\frac{2 \pi i (p-1)}{6}}$ for $p=1, 2, 3, 4, 5, 6$. $\Pi$
stand for high symmetric point $K$, $M$ and $\Gamma$ (See Section I in SI~\cite{SUP}).
The primitive generator of the topologically non-trivial configuration
of our PC is $h_{3c}$ (See Section II in SI \cite{SUP}).

Based on the theory of topological crystalline insulators, we have $[M_1]=2$ and $[K_1]=0$. Therefore 
\begin{equation}
p_i=0,
Q^c=\frac{1}{2}
\end{equation}
We note the dipole moments are always vanishing for $C_{6v}$-symmetric
lattice and the corner topological index is $Q^c=\frac{1}{2}$,
indicating a non-trivial second-order topological insualting phase in our PC.

\textbf{Pseudospins and the bulk-edge-corner correspondence}

Besides the existence of the higher-order topology,
 this staggered couplings between intercell and intracell
sites in the enlarged unit cell which has more internal degree of freedom, introduces an extra pseudospin degree of freedom as
firstly proposed by Wu and Hu in Ref.~\citenum{Hu1}.
Specifically, the enlargement of unit cell yields more
interal degrees of freedom that can be regarded as pesudospins under
the constraint of $C_{6v}$ point group symmetry and time-reversal symmetry.
For example, at the Brillouin zone center,
the unit cell has $p_x (p_y)$ and $d_{xy} (d_{x^2-y^2})$ orbitals (See
Section I in SI.~\cite{SUP}). The linear combinations of these two
set of orbitals $p_{\pm}=(p_x\pm ip_y)/\sqrt{2}$, $d_{\pm}=(d_{xy}\pm
id_{x^2-y^2})/\sqrt{2}$ form pseudospins with pseudo time-reversal
symmetry $T=U\kappa$ as shown in Fig. 1b-c. Here
$U=[D_{E^\prime}(C_6)+D_{E^\prime}(C^2_6)]/\sqrt{3}=-i\sigma_y$ and
$D_{E^\prime}=[1/2, -\sqrt{3}/2; \sqrt{3}/2, 1/2]^T$ represents the
irreducible representations in the $C_6$ symmetry group and $\kappa$
is the complex conjugation operator.
We note that $T^2=-1$ and thus the Kramers degeneracy is realized.
If we combine the topologically non-trivial lattice (TI) with the
topologically trivial lattice (ordinary insulator or OI)[see Fig. 2a
and Section III in SI~\cite{SUP}],
there exist 1D spinful edge states which mimic the 2D quantum spin
Hall effect as shown in Fig. 2b-d. Due to the spin-momentum locking,
the spinful edge states can be selectively excited and propagate unidirectionally along the 1D boundary as shown in Fig. 2c.

Next we study the second-order topological insulator (denoted as SOTI
which shares the same parameters with TI in Fig. 2a) with corner
states by considering a 2D photonic lattice as shown in the lower inset
of Fig. 2e in which the SOTI lattice is surrounded by the OI lattice. The
eigenvalues of TM modes are numerically calculated as shown in
Fig. 2e. We observe a dimensional hierarchy of higher-order topology as
both 1D edge states (represented by yellow dots) and 0D
corner states (represented by blue dots) emerge in between
2D bulk states (represented by gray dots). The corner states here can be regarded as the boundary states of the gapped 1D edge Hamiltonian. We further experimentally measured the transmission spectra as shown in the right panel of Fig. 2e which matches well with the numerical results. The field
distributions of corner states are numerically calculated and shown in
Fig. 2f (lower panel). We find that the out of plane electric fields
are strongly localized at six corners and exponentially decay away
from the corners, demonstrating the 0D geometry. In experiments, we
fabricate this hexagonal photonic lattice by using the alumina rods
arranged in the same parameters and excite the corner state with a
source at frequency $f_c=9.760$GHz. The field distribution is measured
by using the near field scanning technique which matches well with the
simulation as shown in Fig 2f (upper panel) [see other field distributions of corner states in Section IV of SI~\cite{SUP}].

\textbf{Pseudospin-polarized corner states}

Due to extra degrees of freedom of the enlarged unit cell, the
topological corner states also possess the pseudospins of Fig. 1b.
Because of the pseudospin-pseudospin couplings, the pseudospin polarizations of the
corner states are not zero. Therefore they can be selectively
excited by a pseudospin-dependent source. We demonstrate the
pseudospin structures of corner states as shown in Fig. 3. It is
noticed that we
use the half-hexagonal structure here for avoiding the whisper-gallery modes in
a closed boundary as shown in Fig. 3a (upper panel) [see the
tight-binding model which mimics this photonic crystal in Section V of
SI~\cite{SUP}]. The SOTI and OI share the same parameters as those in
Fig. 2a. The pseudospin dependent source is realized by using three
point sources with each one differing by a $2\pi/3$ phase as shown in
Fig. 3a (lower panel). This source has a non-zero orbital angular
momentum (OAM) and is used as the pseudospin dependent source. The numerically calculated eigenmodes are presented in
Fig. 3b. We note that there are two near-degenerate mid-gap states
which are the localized corner states. For those corner states we
display their pseudo-spin textures in Figs. 3c and 3d. For the corner states with lower
frequency at $9.775$GHz, the pseudospin polarization of the left and
right corners are $d_{+}$ and $d_{-}$, respectively (denoted as the
eigenmode $\phi_1$). While for the corner states with upper frequency
at $9.800$GHz,  the pseudospin polarization of the left and right
corners are both $d_{+}$ (denoted as the eigenmode $\phi_2$).

To experimentally observe pseudo-spin polarization of corner states, we put the OAM source in the middle of the 1D boundary between two corners. The simulated and
experimental measured local field intensities are obtained and
depicted in Fig. 4a-d by putting a probe at the left (denoted as A)
corner and the right (denoted as B) as shown in Fig. 3a. We find an
obvious difference in the local field intensity for opposite
pseudospin excitations at two corners as shown in Fig. 4. For left
circular polarized (LCP) excitation we find the A corner has a
relatively higher local field intensity than the B corner while for
right circular polarized (RCP) excitation, the field is mainly
localized at the B corner. We note that there is only one peak for
these two eigenmodes. This is induced by the linear combinations
of previous two eigenmodes $\phi_1$ and $\phi_2$ as $\phi_{LCP}=\phi_1+\phi_2$ and $\phi_{RCP}=\phi_1-\phi_2$. 

For further confirmation of the localization of corner states, we map out the out of plane electric field intensity $|E_z|^2$ by
applying the near field scanning method as shown in Fig. 4e,f. We find
that the field is exponentially localized at the left corner for LCP source
while at the right corner for the RCP source for both the simulations
(white panels) and experiments (black panels). This directional
localization of light at different corners with respect to opposite
pseudospin-dependent sources demonstrates a higher-order quantum spin Hall effect. We note that there is small shift of the
frequency between the simulated and experimental measured data which
are induced by the fabrication errors of our sample.

\textbf{Conclusion}

In conclusion, we theoretically propose and experimentally realize a
spinful higher-order photonic topological insulator in which both 1D
edge states and 0D corner states are observed. Moreover, We achieve a
directional localization of pseudospin polarized corner states of
light, denoted as the higher-order QSHE of
light. Although the existence of the corner states is guaranteed by
the nontrivial higher-order topology, the frequencies and linear combinations of
the corner states are influenced by the finite size effect and quality
of the samples because of their crystalline toplogical nature. The difference in the frequencies between the
simulation and experiment are induced by the finite height of the
dielectric rods in the experiment. Our work paves the way to study
spinful photonic HOTIs and strike advances in lower-dimensional
spintronics and spin photonics. The directional localization of 0D
light enables the designs of topological optical switches, energy
splitters, topological lasings and unidirectional light trappings~\cite{C2, lasing1, lasing2, TRAP}.
Moreover, since our implementation is simple and based
on the all-dielectric material, it can be directly extended to optical
frequencies by using silicon-based coupled resonator optical
waveguides~\cite{PTI3} and femto-lasing direct writing
method~\cite{HOTI10}.
When the quantum optics is considered, the single-photon or multi-photon quantum states may have anomalous quantum behaviors which
are related to the pseudospin degree of freedom such as the spinful
quantum walk, spinful quantum correlations and
entanglements~\cite{quan1,quan2,quan3,quan4}. If we couple our system to quantum emitters, the spinful corner states may support chiral emission of photons and potential applications in quantum simulations~\cite{emitter}.
Although our implementation are based
on photonic crystals, we expect to observe a similar effect in other
classical wave systems such as phononic
crystals~\cite{PNI1,PNI2,PNI3,PNI4},
surface polariton plasmon, and electric circuits~\cite{HOTI4,E2}.


\textbf{\large{Method}}

\textbf{{Design of the second-order spinful photonic topological insulator}} We design our 2D photonic crystals by using all-dielectric cylinders consist of alumina ceramics. The cylinders are stuck to the metallic plates with a depth of $1.5mm$. The six dielectric rods in a unit cell support multipole resonant modes at different frequencies such as the monopole, dipole, quadrupole and so on, mimicing the atomic orbitals. We use the absorbing materials to surround our sample to reduce the wave reflections and background noise. We note that there are two differences between our design and the corresponding tight-binding model. First, in the low frequency limit, our photonic crystals can be regarded as homogeneous materials and the dispersion relation is linear which naturally breaks the chiral symmetry (sublattice symmetry). This leads to a frequency shift of all states (bulk, edge, corner states) comparing to the tight-binding case. Second, the pseudospins in our implementation is defined as a continuous configuration (state) of in-plane magnetic field while this is impossible for the tight-binding model.

\textbf{{Numerical simulations}} We use commercial software: COMSOL MULTIPHYSICS
to conduct all the numerical simulations of our samples. In all
simulations, we use a 2D photonic crystals to mimic our sample. This
is valid when we consider the TM modes. For the boundary conditions of
the ribbon structure we used to obtain Fig. 2b, we set the in-plane
boundary parallel to the propagation direction as the perfect electric
conductor (PEC) and the boundary perpendicular to the propagation
direction as Floquet periodic boundary (See section III in.~\citenum{SUP}). In simulating Fig. 2e,f, Fig. 3b and Fig. 4e,f,
we use the scattering boundary condition for the boundaries parallel to the interface of two PCs.

In simulating Fig. 3b, we find extra mid-gap states which are upper
boundary states and irrelevant to our study. Therefore, we use an
algorithm to eliminate those irrelevant states as follows: We take the
upper boundary of the nontrivial region as an area of $18a \times
2.5l$, where $a$ is the lattice constant and $l=a/\sqrt{3}$ is the
edge length of the unit cell.  By defining the proportion of the
boundary field in the whole region, the mode independent of the
boundary can be selected. The filter condition is defined as
$\int_{edge}|E_z|^2ds/\int_{all}|E_z|^2ds<0.5$. This ensures that in
the eigenfrequency of Fig. 3b in the main text,
the edge mode at the upper boundary is naturally eliminated. For the calculation of pseudospins in Fig. 3b, we first obtain the out of plane electric fields and then directly calculate the corresponding in plane magnetic fields.
\noindent

\textbf{The pseudospin-dependent source}
In our implementation, we use an OAM source as the pseudospin-dependent source. To show the overlap between the source excitation and pseudospin polarization, we numerically calculate the overlap integrals defined as $|\braket{d_{\pm}|\phi}|$. Here $d_{\pm}$ is the pseudospin eigenstates and $\phi$ is the OAM field in a unit cell. For a source with clockwise/anti-clockwise circular polarization at 9.755GHz, we extract the distribution of excitation field over a unit cell as $\phi_{R/L}$. We then discrete the continuous field and obtain the overlap between the source and the pseudospin eigenstates as $SP_{\pm,R/L}=|\braket{d_{\pm}|\phi_{R/L}}|^2/Max(|\braket{d_{+}|\phi_{R/L}}|^2,|\braket{d_{-}|\phi_{R/L}}|^2)$. Numerical calculations show that $SP_{+,R}=1.63\times10^{-5}$, $SP_{-,L}=1$ for RCP source and $SP_{+,R}=1$, $SP_{-,L}=1.63\times10^{-5}$ for LCP source, which reveals the pseudospin-dependent character of the source. 

\noindent
\textbf{Experiments} In our experiments, the microwave near-field measurement system mainly consists of two parts: the vector network analyzer (Agilent E5063A) and the 3D near-field scanning platform. Three microwave antennas (with $0$, $2\pi/3$, $4\pi/3$ degree phase delay ) is mounted on the bottom of the aluminum plate through drilled holes. When the microwave probe antenna moves horizontally driven by a stepper motor (with 2 mm step size), the $z$ component of the electric field of the TM modes in the frequency range of our interest (covering from 9.000 GHz to 10.500 GHz) can be measured. To map the measured electric field distribution at selected frequencies more clearly, we define $(E_z/E_0)^2$ as variate to the plot, which is processed by MATLAB.

\begin{addendum}
 \item[Acknowledgements] This work was financially supported by National Key
R\&D Program of China (Grants No. 2018YFA0306202 and No. 2017YFA0303702) and National Natural Science Foundation of China (Grants No. 11625418, No. 51732006, No. 11674166, and No. 11774162). 
  
 \item[Author contributions] M.H.L., P.Z. and B.X. conceived the idea. G.S. and L.H. did the experiments. H.F.W. and G.S. performed the numerical simulations. B.X. G.S. and H.F.W. did the theoretical analyses and data analyses. Z.W. and Y.F.C. guided the research. All authors contributed to discussions of the project. B.X., H.F.W., G.S., P.Z. and M.H.L. prepared the manuscript with revisions from other authors.

 \item[Competing Interests] The authors declare that they have no
competing financial interests.
\end{addendum}



\pagebreak
\begin{figure}[htbp]
\begin{center}
\includegraphics[width=0.65\columnwidth]{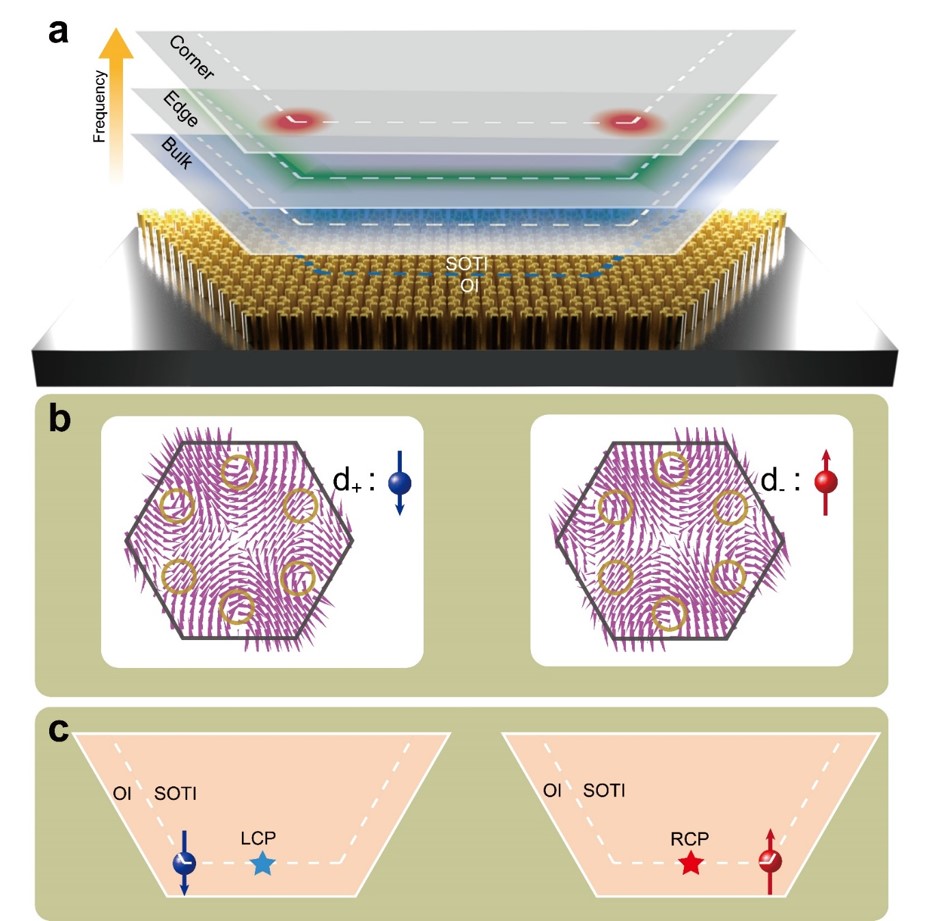}
\label{fig:1}
\end{center}
\end{figure}
\noindent
\textbf{Figure 1 $|$ Higher-order quantum spin Hall effect of light in 2D second-order photonic topological insulator} \textbf{a} The dimensional hierarchy of higher-order topological insulators in an all-dielectric photonic crystals. The corner, edge and bulk states are represented by blue, green and red colors respectively and separated from each other in the frequency domain. \textbf{b} Two pseudospins defined by the in-plane magnetic field (represented by the purple arrows) configuration in the unit cell. \textbf{c} The scheme of achieving directional localization of psedospin-polarized corner states excited by a pseudospin-dependent source. The blue (red) star represents the position of a left-circular (right-circular) polarized light as a pseudospin-dependent source. The blue (red) sphere represents the position of the corner states with a pseudospin up (down) polarization (represented by the arrows).

\pagebreak
\begin{figure}[htbp]
\begin{center}
\includegraphics[width=0.9\columnwidth]{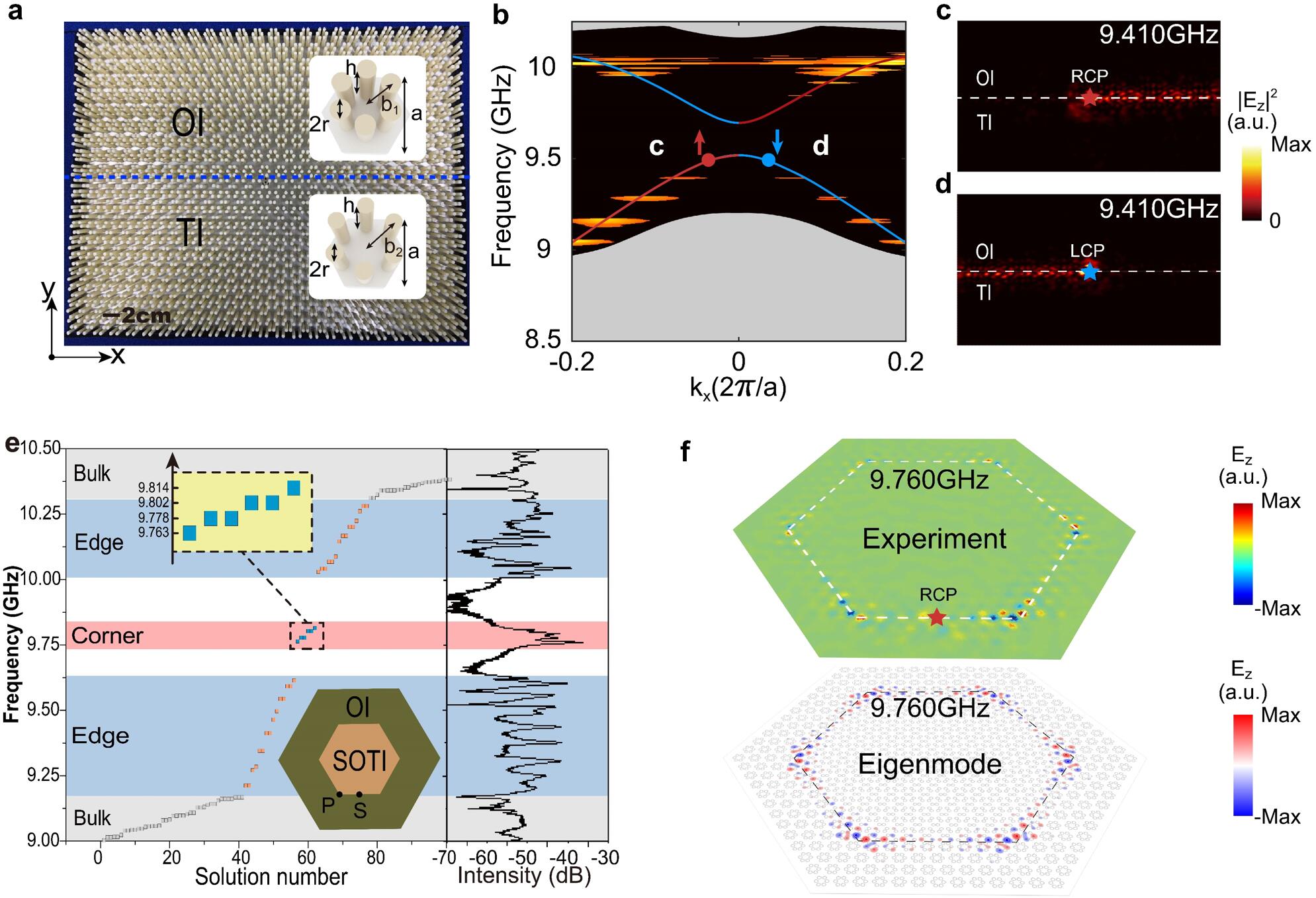}
\label{fig:1}
\end{center}
\end{figure}
\noindent
\textbf{Figure 2 $|$ Emergence of multi-dimensional boundary states at the interface between two topologically distinct photonic crystals} The \textbf{a} combination of an ordinary insulator and a topological insulator with a line-shape boundary (blue dashed line).\textbf{b} The measured (yellow area) and the simulated (solid line) spinful edge states in the projected band structure. The unidirectional propagation of light to the \textbf{c} right and \textbf{d} left directions with an RCP and an LCP source respectively.  \textbf{e} The numerically calculated eigenvalues (left panel) and measured transmission (right panel) of the  hexagonal photonic lattice (lower inset). The bulk, edge and corner states are represented by grey, blue and yellow dots respectively. \textbf{f} The field distribution of the numerically calculated (lower panel) and experimentally excited (upper) corner states at 9.760 GHz.

\pagebreak
\begin{figure}[htbp]
\begin{center}
\includegraphics[width=0.9\columnwidth]{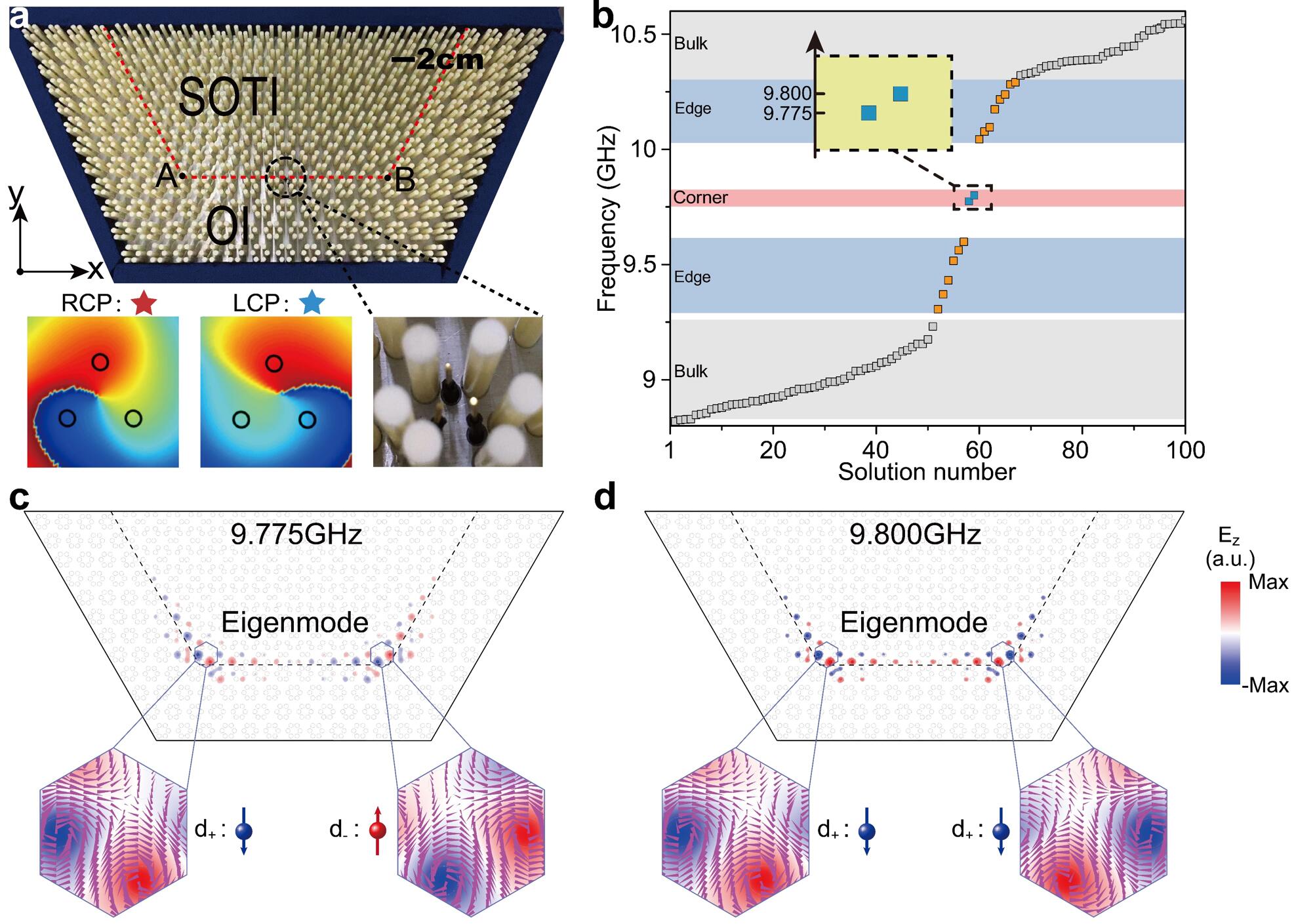}
\label{fig:1}
\end{center}
\end{figure}
\noindent
\textbf{Figure 3 $|$ The pseudospin-polarized corner states.} \textbf{a} The half-hexagonal photonic lattice consists of an SOTI and an OI (upper panel). The pseudospin dependent source are realized by three point sources with each phase differing by $2\pi/3$. The experimental measured field distribution of the source exhibit OAM fields (lower panel). The \textbf{b} calculated eigenmodes of the half-hexagonal photonic lattice. \textbf{c} The pseudospin polarizations represented by the configurations of the in-plane magnetic field (purple arrows) at two corners are opposite and the same for \textbf{c} the lower frequency mode and \textbf{d} the higher frequency mode respectively. 

\pagebreak
\begin{figure}[htbp]
\begin{center}
\includegraphics[width=0.9\columnwidth]{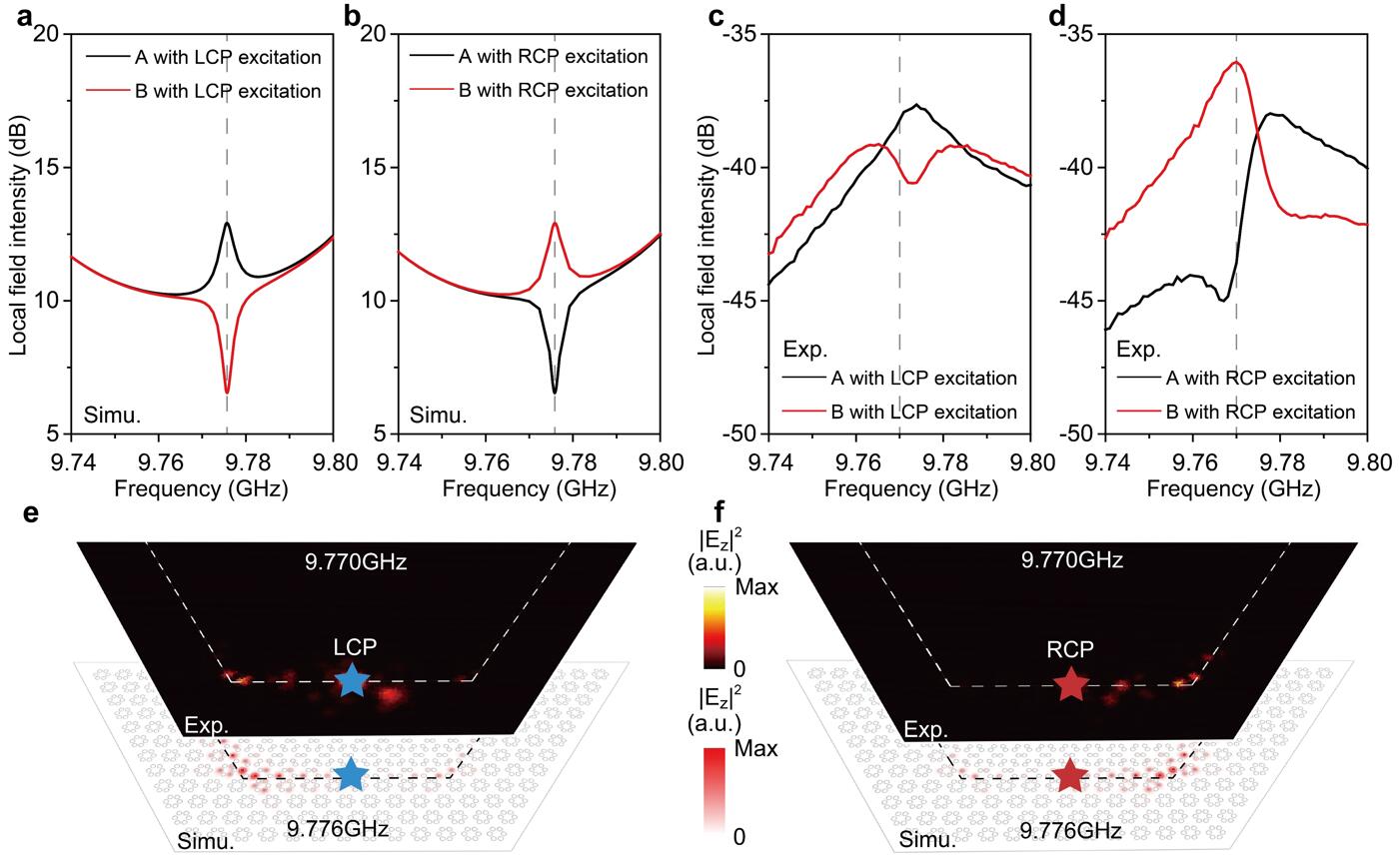}
\label{fig:1}
\end{center}
\end{figure}
\noindent
\textbf{Figure 4 $|$ Directional localization of light at corners} The \textbf{a} (\textbf{b}) simulated and \textbf{c} (\textbf{d}) experimentally measured transmission spectra excited by an LCP (RCP) source located at the center of the 1D boundary respectively. 
\textbf{e} The simulated (white panel with excitation source at $9.776$GHz) and experimentally measured (black panel with excitation source at $9.770$GHz) $|E_z|^2$ distribution for a LCP excitation. \textbf{f} The simulated (white panel with excitation source at $9.776$GHz) and experimentally measured (black panel with a excitation source at $9.770$GHz) $|E_z|^2$ distribution for a RCP excitation. 

\end{document}